\documentstyle[12pt]{article}
\def\e{{\rm e}}
\newcommand{\be}{\begin{equation}}
\newcommand{\ee}{\end{equation}}
\newcommand{\bea}{\begin{eqnarray}}
\newcommand{\eea}{\end{eqnarray}}

\begin{document}
\begin{titlepage}
\rightline{hep-th/9410183}
\rightline{MPI-PhT/94-66}
\rightline{ZU-TH 32/94}
\rightline{October 1994}
\begin{center}
{\bf BLACK HOLES AND SPHALERONS IN LOW \\
 ENERGY EFFECTIVE STRING THEORY
\footnote{Talk given at the 7th Marcel Grossmann Meeting on
General Relativity, Stanford  University, July 24-30, 1994.
To be published in the Proceedings.}
}
\vglue 1.0cm
{GEORGE LAVRELASHVILI
 \footnote{Electronic address: lavrela@physik.unizh.ch}
 \footnote{On leave of absence from Tbilisi
Mathematical Institute, 380093 Tbilisi, Georgia}
 }\\
\baselineskip=14pt
\vglue 0.3cm
{\it Institut f\"ur Theoretische Physik, Universit\"at Z\"urich}\\
{\it Winterthurerstrasse 190, CH-8057 Z\"urich, Switzerland
\footnote{Address after October 1, 1994.}
}\\
\vglue 0.3cm {and}\\
\vglue 0.3cm
{\it Max-Planck-Institut f\"ur Physik, Werner-Heisenberg-Institut}\\
{\it F\"ohringer Ring 6, D-80805 Munich, Germany}
\vglue 0.8cm
{ABSTRACT}
\end{center}
\vglue 0.3cm
{\rightskip=3pc
 \leftskip=3pc
 \tenrm\baselineskip=12pt
\noindent
We discuss properties of globally regular and black hole
solutions of the Einstein-Yang-Mills-dilaton theory.
}
\end{titlepage}
\section*{}
The existence of a dilaton - a neutral,
scalar field with exponential coupling to matter
is one of the universal predictions following from various
types of string theories and from Kaluza-Klein theories
(see e.g. \cite{GSW}). The dilaton field has an influence on
the cosmological evolution of the Universe
\cite{BER}
and it might modify particle-like
solutions of corresponding theories.

In the present report we are interested in the second aspect of
the problem, namely we will study static spherically symmetric
solutions of the Einstein-Yang-Mills-dilaton (EYMD) theory.
We will consider the EYMD theory defined by the action
\be
S=
\frac{1}{4\pi}\int\Bigl(-\frac{1}{4 G}R+
\frac{1}{2}(\partial\varphi)^2-
 {{\rm e}^{2\kappa\varphi}\over 4g^2}F^2\Bigr)
 \sqrt{-g} d^4x\; , \label{act}
\ee
where $G$ is Newton's constant,
$g$ is the gauge coupling constant  and
$\kappa$ denotes the dilatonic coupling constant.
The scaling properties of the action
Eq.(\ref{act}) allow us to put  $G=g=1$ without restrictions
in what follows.
The only remaining free parameter in the action is the dimensionless
ratio $\gamma=\kappa /g\sqrt{G}$.

Note that in the limit $\gamma\to 0$ the model described by
Eq.~(\ref{act}) reduces to the
Einstein-Yang-Mills theory \cite{BM,BH}.
In the limit $\gamma\to\infty$ one gets
the  Yang-Mills-dilaton theory \cite{LM1} in flat
space.

To study static, spherically symmetric solutions of
EYMD theory
a convenient parametrization for the metric turns out to be
\be
ds^2=A^2(r)\mu(r)dt^2-{dr^2\over\mu(r)}
  -r^2d\Omega^2\;, \label{interval}
\ee
where $d\Omega^2=d\theta^2+sin^2(\theta)d\varphi^2$
is the line element of the unit sphere.

For the $SU(2)$ Yang-Mills potential we make the usual (`magnetic')
spherically symmetric ansatz
\be
W_0^a=0,\quad
  W_i^a=\epsilon_{aik}{x^k\over r^2}(W(r)-1)\;. \label{gauge}
\ee

Substituting this ansatz into the action
we obtain the reduced action
\be
S_{\rm red}=-\int A\Bigl[{1\over 2}(\mu+r\mu'-1)+
  {r^2\over 2}\mu\varphi'^2+\e^{2\gamma\varphi}
  \Bigl(\mu W'^2+{(1-W^2)^2\over2r^2}\Bigr)\Bigr]\,dr\;,\label{redact}
\ee
where a prime denotes $d\over dr$.

The resulting field equations are
\bea
 (A\e^{2\gamma\varphi}\mu W')'&=&A\e^{2\gamma\varphi}
   {W(W^2-1)\over r^2} \;, \nonumber \\
 (A\mu r^2\varphi')'&=&
 2\gamma A\e^{2\gamma\varphi}\Bigl(\mu W'^2+{(1-W^2)^2\over2r^2}
 \Bigr)\;,  \nonumber \\
 \mu'&=&{1\over r}\Biggl(1-\mu-r^2\mu\varphi'^2-2\e^{2\gamma\varphi}
 \Bigl(\mu W'^2+{(1-W^2)^2\over2r^2}\Bigr)\Biggr)\;, \nonumber  \\
 A^{-1}A'&=&\frac{2\e^{2\gamma\varphi}W'^2}{r}
 +r\varphi'^2\;.\label{eqm}
\eea

The field equations Eq.~(\ref{eqm})  have
singular points at $r=0$ and $r=\infty$ as well
as at points where $\mu(r)$ vanishes.

Inserting a power series expansion into Eq.~(\ref{eqm}) one
finds a $2-$parameter family of regular solutions in the
vicinity of  $r=0$
\bea
W(r)&=&1-br^2+O(r^4)\;,~~~~~~~~~~~~~~
\mu(r) = 1-4b^2\e^{2\gamma\varphi_0}r^2+O(r^4)\;, \nonumber \\
\varphi(r)&=&\varphi_0+2\gamma \e^{2\gamma\varphi_0}b^2r^2
      +O(r^4)\;,\;\;
A(r) = 1+4b^2\e^{2\gamma\varphi_0}r^2+O(r^4)\;,
\label{ezero}
\eea
where $b$ and $\varphi_0$ are arbitrary parameters.

Similarly at $r=\infty$ one finds
\bea
 W(r)&=&\pm(1-{c\over r}+O({1\over r^2}))\;,\;\;
 \mu(r) = 1-{2M\over r}+O({1\over r^2})\;, \nonumber \\
 \varphi(r)&=&\varphi_\infty-{d\over r}+O({1\over r^2})\;,\;\;~~
 A(r) = A_\infty(1-{d^2\over2r^2}+O({1\over r^4}))\;,
\label{einfty}
\eea
where again $c,d,M,\varphi_\infty$ and $A_\infty$ are arbitrary
parameters.

In the vicinity of the singular point $r_h$ where $\mu (r)$ vanishes
one finds a $2-$parame\-ter family of solutions which stay regular
at this point.
Under an appropriate choice of the parameters the surface $r=r_h$
corresponds to a regular event horizon.

Globally regular asymptotically flat
(respectively back hole) solutions have to
interpolate between the described asymptotic behavior at $r=0$
(respectively $r=r_h$) and $r=\infty$.

The analysis \cite{LM2,BIZ2} for $0<\gamma<\infty$ yields what could be
expected from the extreme cases $\gamma =0$ and $\gamma =\infty$.
It was found \cite{LM2} that for any value of the dilaton coupling
constant $\gamma$ the EYMD system has:
a discrete family of globally regular (sphaleron type)
solutions of finite mass and a discrete family of (non-Abelian)
black hole solutions.
The solutions in each family are labeled by the
number $n$ of zeros of the gauge field potential $W(r)$.

The solutions are found to be unstable \cite{LM2}.

It is interesting that the black holes have a non-vanishing
Yang-Mills field outside the horizon.
One can interpret the solutions with the horizon as
``black holes inside sphalerons''.

The mass of the solutions in the case of the Einstein-Yang-Mills
theory is of the order of one in natural units $M_{EYM}=1/g\sqrt{G}$.
In the EYMD theory the mass of the solutions decreases with increasing
dilatonic coupling constant $\gamma$ and
for large $\gamma$ goes to zero like
$M_{EYMD}\sim \frac{M_{EYM}}{\gamma ^2}$.

Globally regular solutions with odd $n$ are kinds of sphalerons.
One finds fermion zero modes in the background of these solutions
\cite{GIB,VOLF,LM3}.
One can assign a topological number to these solutions.
It is half integer.

There are some interesting regularities in the EYMD theory.
It turns out that
%\cite{FOR}
the ${tt}$ component of the metric is related to the dilaton
$A^2 \mu = e^{2 \gamma \varphi}$ and as a consequence
of this relation the dilaton charge
(parameter $d$ in Eq.(\ref{einfty}))
is equal to the mass of the solution: $d=M$.

A very special situation occurs for the value of
the dilaton coupling constant $\gamma=1$, which corresponds to
the model obtained from heterotic string theory.
It was found \cite{LM2} that for the $n=1$ solution
the parameter $b$ is a rational number, $b =\frac{1}{6}$.
Another regularity found numerically is that an asymtotic coefficient
$c$ in Eq.(\ref{einfty}) is related to the mass of the solution, $c=2M$.
We think these are arguments indicating that the lowest lying
($n=1$) regular solution may be obtained in closed form.
similarly to the ``stringy instanton''
and the ``stringy monopole''\cite{INSTMON}.

Due to the high mass of the solutions the only situation where they
could play a role is in the Early Universe, but at the moment there
seems to be no natural physical scenario where we could make use of
these solutions.

\section*{Acknowledgements}

I am grateful to Dieter Maison for a fruitful collaboration
which led to the results reported in this paper.


\begin{thebibliography}{9}

\bibitem{GSW}
M.B. Green, J.H. Schwarz and E. Witten,
{\it Superstring Theory\/} (Cambridge U.P., Cambridge, 1987).

\bibitem{BER}
See e.g.,\\
O. Bertolami, Talk presented at the MG7;\\
M.C. Bento and O. Bertolami,
{\it Phys. Lett.}  {\bf B336} (1994) 6.

\bibitem{BM}
 R. Bartnik and J. McKinnon,
 {\it Phys.\ Rev.\ Lett.\/} {\bf 61} (1988) 141.

\bibitem{BH}
 M.S. Volkov and D.V. Gal'tsov,
{\it  JETP\ Lett.\/} {\bf 50} (1990) 346;\\
{}~~H.P. K\"unzle and A.K.M. Masood-ul-Alam,
{\it J.\ Math.\ Phys.\/} {\bf 31} (1990) 928;\\
{}~~P. Bizon,
{\it Phys.\ Rev.\ Lett.\/ } {\bf 61} (1990) 2844.

\bibitem{LM1}
 G. Lavrelashvili and D. Maison,
 {\it Phys.\ Lett.\/} {\bf B295} (1992) 67; \\
 P. Bizon,
 {\it Phys.\ Rev.\/} {\bf D47} (1993) 1656.

\bibitem{LM2}
 G. Lavrelashvili and D. Maison,
 {\it Nucl.\ Phys.\/} {\bf B410}  (1993) 407;\\
 G. Lavrelashvili and D. Maison,
 {\it Dilatonic sphalerons and non-abelian black holes,}
 Preprint MPI-Ph/93-55, hep-th/9307159.

\bibitem{BIZ2}
 E.E. Donets and D.V. Gal'tsov,
 {\it Phys.\ Lett.\/} {\bf B302} (1993) 411;\\
 P. Bizon,
 {\it Acta Physica Polonica\/} {\bf B24} (1993) 1209;\\
 T. Torii and Kei-ichi Maeda,
 {\it Phys.\ Rev.\/} {\bf D48} (1993) 1643.


\bibitem{GIB}
 G.W. Gibbons and A. Steif,
 {\it Phys.\ Lett.\/} {\bf B314} (1993) 13.

\bibitem{VOLF}
 M.S. Volkov,
 {\it Phys.\ Lett.\/} {\bf B334} (1994) 40.

\bibitem{LM3}
 G. Lavrelashvili,
{\it Fermions in the background of dilatonic sphalerons},\\
 hep-th/9410178.

%\bibitem{FOR}
% P. Forgacs,\\
% {\it to be published} (1994).

\bibitem{INSTMON}
C.G. Callan, J.A. Harvey and A. Strominger,
{\it Nucl.\ Phys.\/} {\bf B359} (1991) 611;\\
J.A. Harvey and J.Liu,
{\it Phys.\ Lett.\/} {\bf B268} (1991) 40.
%
\end{thebibliography}
\end{document}